\documentclass[prl,twocolumn,superscriptaddress,showpacs]{revtex4}

\usepackage{graphicx,amsfonts,amsmath}
\usepackage{dcolumn}
\usepackage{bm}

\newcommand{\bk}{b_{\bm{k}}}
\newcommand{\bkd}{b_{\bm{k}}^{\dag}}
\newcommand{\bmk}{b_{-\bm{k}}}

\newcommand{\sumk}{\sum_{\bm{k}}}
\newcommand{\wk}{\omega_{\bm{k}}}
\newcommand{\fk}{f_{\bm{k}}}
\newcommand{\gk}{g_{\bm{k}}}
\newcommand{\pe}{|1\rangle\!\langle 1|}

\newcommand{\unit}{\mathbb{I}}
\DeclareMathOperator{\Tr}{Tr}
\DeclareMathOperator{\re}{Re}

\begin{document}


\title{Anomalous decay of quantum correlations of quantum dot qubits}

\author{Katarzyna Roszak}
\affiliation{Institute of Physics, Wroc{\l}aw University of Technology,
50-370 Wroc{\l}aw, Poland}

\author{Pawe{\l} Mazurek}
\affiliation{National Quantum Information Centre of Gda{\'n}sk, 81-824 Sopot, Poland}
\affiliation{Institute for Theoretical Physics and Astrophysics,
University of Gda{\'n}sk, 80-952 Gda{\'n}sk, Poland}

\author{Pawe{\l} Horodecki}
\affiliation{National Quantum Information Centre of Gda{\'n}sk, 81-824 Sopot, Poland}
\affiliation{Faculty of Applied Physics and Mathematics, Gda{\'n}sk University of Technology, 
80-952 Gda{\'n}sk, Poland}

\date{\today}

\begin{abstract}
We study the evolution of quantum correlations, quantified by the geometric discord,
of two excitonic quantum dot qubits under the influence of the phonon environment.
We show that the decay of these correlations differs substantially form the decay
of entanglement. Instead of displaying sudden death type behavior, the geometric discord
shows a tendency to undergo transitions between different types of decay, is sensitive to 
non-local phase factors, and may already be enhanced by weak environment-mediated interactions.
Hence, two-qubit quantum correlations are more robust under decoherence processes,
while showing a richer and more complex spectrum of behavior under unitary and non-unitary
evolution.
\end{abstract}

\maketitle

The study of quantum correlations in realistic systems has, for a long time, been limited 
to the study of entanglement, due to the fact that straightforward methods of calculating
the amount of correlations in a two-qubit system 
have only been available for some entanglement measures, such as the concurrence \cite{hill97,wootters98}
or negativity \cite{zyczkowski98,vidal02}.
Although entanglement itself is a very important resource for a number of applications
\cite {horodecki09}, including
quantum computation, quantum cryptography, or teleportation, separability (the lack of entanglement)
does not automatically exclude the presence of quantum correlations \cite{modi12}. 
This is in particular the reason why quantum computation models
relying on mixed, separable (not entangled) states \cite{knill98,meyer00,passante11} are possible.

The quantum discord \cite{olliver01,henderson01} is a measure of quantum correlations (see, however 
Refs.~\cite{henderson01,oppenheim02} for Holevo-type and thermodynamic based measures)
which captures correlations beyond entanglement; 
it is defined as the difference 
of two classically equivalent formulas for mutual information 
and is non-negative. 
Due to the null volume of the set of zero-discord states \cite{ferraro10}, discord measures
are not expected to undergo
sudden death which is characteristic for entanglement evolutions \cite{zyczkowski01,yu04,eberly07}.
The geometric measure of the discord describes the amount
of correlations in a quantum system by finding the minimal Hilbert-Schmidt distance
to the set of zero-discord states \cite{dakic10}.
Recently, a lower \cite{dakic10} and an upper \cite{miranowicz12} bound on the geometric discord 
which can be calculated from a two-qubit
density matrix have been found, which substantially simplifies the problem of studying
the evolution of the quantum discord and opens the path for a qualitative and quantitative
description of the decay
of quantum correlations in realistic open quantum systems.

In this paper we study the evolution of the lower and upper bounds of the geometric discord
of two exciton quantum dot (QD) qubits interacting with an open phonon environment
in order to capture the physical aspects of decoherence effects on quantum correlations.
The interactions present in the system and the resulting dynamics are well understood.
The experimentally observed evolution on picosecond timescales \cite{borri01,vagov04} 
can be described by pure dephasing within
the independent boson model \cite{vagov03,vagov04}. Super-Ohmic phonon spectral densities 
\cite{krummheuer02,alicki04b} 
(resulting from the actual form of the carrier-phonon coupling and the phonon density of states
\cite{mahan00}) are responsible
for characteristic features of the dephasing which is non-exponential and always only partial.
Furthermore, a finite distance between the QDs leads to a time-delayed interference of phonon
wave packets traveling from the two QDs which induces an environment-mediated interaction
between the dots (and a small enhancement of the density matrix coherences) in addition
to the exciton-exciton interaction present in the system.
The fact that the complex evolution of this ensemble can be credibly described theoretically
in combination with experimental accessibility to a wide range of pure initial states
(which are optically excited on femtosecond timescales)
make this system ideal for the examination of the quantum-information properties of
open quantum systems.  

The specific system under study consists of two QDs stacked on top of each other
and interacting with a phonon reservoir.
The single qubit states $|0\rangle$ and $|1\rangle$ correspond to an empty QD
and an exciton excited in the dot, respectively. 
The system is described by the Hamiltonian 
\begin{eqnarray}\label{ham0}
H & = & \epsilon_{1}(\pe\otimes\unit)
+\epsilon_{2}(\unit\otimes\pe)
+\Delta\epsilon(\pe\otimes\pe) \nonumber \\
&&+(\pe\otimes\unit)\sumk\fk^{(1)}(\bkd+\bmk) \\ \nonumber
&&+(\unit\otimes\pe)\sumk\fk^{(2)}(\bkd+\bmk)
+\sumk\wk\bkd\bk ,
\end{eqnarray}
where $\mathbb{I}$ is the unit operator,
$\epsilon_{1,2}$ are the transition energies
in the two subsystems, $\Delta\epsilon$ is the biexcitonic shift
due to the interaction between the subsystems, $\fk^{(1,2)}$
are system-reservoir coupling constants, $\bk,\bkd$ are bosonic
operators of the reservoir modes, and $\wk$ are the corresponding energies
(we put $\hbar=1$). 

Exciton wave functions are modeled by anisotropic Gaussians
with the extension $l_{\perp}$ in the $xy$ plane
and $l_{z}$
along $z$ for the electron and hole in both dots.
The coupling constants for the deformation potential coupling between
confined charges and longitudinal phonon modes have the form
$\fk^{(1,2)}=\fk e^{\pm ik_{z}d/2}$, where $d$ is the distance between the dots and
\begin{displaymath}
\fk=\sqrt{\frac{k}{2\varrho Vc}}(\sigma_{\mathrm{e}}-\sigma_{\mathrm{h}})
e^{-l_{z}^{2}k_{z}^{2}/4}
e^{-l_{\mathrm{e}}^{2}k_{\bot}^{2}/4},
\end{displaymath}
where $V$ is the normalization volume of the bosonic reservoir, 
$k_{\bot},z$ are momentum
components in the $xy$ plane and along the $z$ axis,
$\sigma_{\mathrm{e,h}}$ are deformation potential constants for
electrons and holes, $c$ is the speed of longitudinal sound,
and $\varrho$ is the crystal density.
In our calculations we put $\sigma_{\mathrm{e}}=8$ eV,
$\sigma_{\mathrm{h}}=-1$ eV, $c=5.1$ nm/ps, $\varrho=5360$ kg/m$^{3}$
(corresponding to GaAs), 
$l_{\perp}=5$ nm, and $l_{z}=1$ nm.
The distance between the dots is taken equal to $d=6$ nm unless stated otherwise.

The Hamiltonian (\ref{ham0}) can be diagonalized exactly
using the Weyl operator method \cite{mahan00,roszak06b} and we find the evolution of the double QD
subsystem following Ref.~[\onlinecite{roszak06a}].
Since local unitary transformations do not change the amount of
quantum correlations in the system, we can use the density matrix $\tilde{\rho}(t)
=e^{-iH_L t}\rho(t)e^{iH_L t}$, with $H_L=E_{1}(|1\rangle\!\langle 1|\otimes\mathbb{I})
+E_{2}(\mathbb{I}\otimes|1\rangle\!\langle 1|)$, 
where $E_{i}=\epsilon_{i}-\sumk|\fk|^{2}/\wk$
instead
of $\rho(t)$ in the study of the geometric discord.
Assuming a separable initial
system-reservoir state, we find
the evolution of the elements of the density matrix $\tilde{\rho}(t)$, which
are equal to
\begin{equation}\label{ent:mat-elem}
\left[\tilde{\rho}(t)\right]_{ii} = [\tilde{\rho}_{0}]_{ii};\;\;\;
\left[\tilde{\rho}(t)\right]_{ij} = [\tilde{\rho}_{0}]_{ij}
e^{-iA_{ij}(t)+B_{ij}(t)},
\end{equation}
with
\begin{subequations}
\begin{eqnarray}
A_{01}(t)&=&A_{02} = \sum|g_{\bm{k}}|^{2}\sin \wk t,\\
A_{03}(t) &=& 4\sum|g_{\bm{k}}|^{2}\cos^{2}(k_{z}d/2 )\sin \wk t-\Delta Et,\\
A_{12}(t)&=&0,\\
A_{13}(t)&=&A_{23}=A_{03}-A_{01},\\
B_{01}(t)&=&B_{02}=B_{13}=B_{23}\\
\nonumber
&=&\sum|g_{\bm{k}}|^{2}(\cos \wk t-1)(2n_{\bm{k}}+1),\\
\label{B03}
B_{03}(t)&=&4\sum|g_{\bm{k}}|^{2}\cos^{2}(k_{z}d/2 )(\cos \wk t-1)\\
\nonumber
&&\times(2n_{\bm{k}}+1),\\
\label{B12}
B_{12}(t)&=&  4B_{01}-B_{03},
\end{eqnarray}
\end{subequations}
where $n_{\bm{k}}$ is the Bose distribution, 
$g_{\bm{k}}=\fk/\wk$, $\Delta E=\Delta\epsilon-2\re\sumk\wk|\gk|^2e^{ik_{z}d}$, 
and the indices $i,j=0,1,2,3$ correspond to the two qubit states
$|00\rangle$, $|01\rangle$, $|10\rangle$, and $|11\rangle$, respectively.
For long times, the factors $\cos \wk t$ and $\sin \wk t$ 
become quickly oscillating
functions of $\bm{k}$ and their contribution averages to 0.
Consequently, the phase damping factors $B_{ij}$ decrease form their initial
value of 0 to a certain asymptotic value depending on the material
parameters, system geometry and temperature, while the phase shift factors $A_{ij}$ 
affect the system evolution at small times and then average out to zero. As a result, the
off-diagonal elements of the density matrix are reduced and the phase
information is partly erased. 

Here, we are interested in the symmetric geometric discord \cite{dakic10} which may also 
be expressed  in the form of a purity deficit (see Ref.~\cite{miranowicz12})
\begin{equation}
D_{S}(\rho_{AB})=\min_{{\cal M}_{A} \otimes {\cal M}_{B}}  \left(\Tr[\rho_{AB}^{2}]
- \Tr[({\cal M}_{A} \otimes {\cal M}_{B}) \rho_{AB}]\right).
\end{equation}
Specifically, the discord is formulated as a purity (quadratic Renyi entropy) deficit under global versus 
product local  (${\cal M}_{A} \otimes {\cal M}_{B}$) von Neumann measurements. 
In case of two qubits, the lower bound on the discord
is given by \cite{dakic10}
\begin{equation}
\label{lower}
D_S'=\max\left(
\Tr[K_x]-k_x,\Tr[K_y]-k_y
\right),
\end{equation}
where $k_x$ is the maximum eigenvalue of the matrix $K_x=|x\rangle\langle x|+TT^T$
and $k_y$ is the maximum eigenvalue of the matrix $K_y=|y\rangle\langle y|+T^T T$.
Here, $|x\rangle$ and $|y\rangle$
denote local Bloch vectors with components $x_i=\Tr[\rho_{AB}(\sigma_i\otimes\mathbb{I})]$
and $y_i=\Tr[\rho_{AB}(\mathbb{I}\otimes\sigma_i)]$, and the elements of the correlation
matrix $T$ are given by $T_{i,j}=\Tr[\rho_{AB}(\sigma_i\otimes\sigma_j)]$ (stemming from the
standard Bloch representation of a two-qubit density matrix $\rho_{AB}$).
The upper bound is given by \cite{miranowicz12}
\begin{eqnarray}
\label{upper}
D_S''&=&\min\left(
\Tr[K_x]-k_x+\Tr[L_y]-l_y,\right.\\
\nonumber
&&
\left.\Tr[K_y]-k_y+\Tr[L_x]-l_x
\right),
\end{eqnarray}
where $l_x$ and $l_y$ are the maximal eigenvalues of the matrices 
$L_x=|x\rangle\langle x|+T|\hat{k}_y\rangle\langle \hat{k}_y|T^T$ and
$L_y=|y\rangle\langle y|+T^T|\hat{k}_x\rangle\langle \hat{k}_x|T$, respectively,
while $|\hat{k}_x\rangle$ and $|\hat{k}_y\rangle$ are the normalized eigenvectors
corresponding to the eigenvalue $k_x$ of matrix $K_x$ and $k_y$ of matrix $K_y$.
In the case of symmetric two-qubit states, meaning $\rho_{AB}=\rho_{BA}$, no minimization
or maximization is needed in eqs. (\ref{lower}) and (\ref{upper}).
\begin{figure}[th]
\begin{center}
\unitlength 1mm
\begin{picture}(85,35)(0,5)
\put(0,0){\resizebox{85mm}{!}{\includegraphics{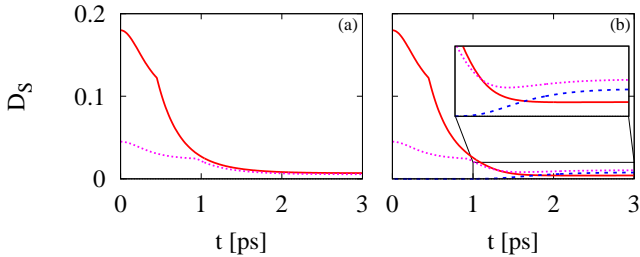}}}
\end{picture}
\end{center}
\caption{\label{zgrzyt_dziwny2} Evolution of the X-state geometric discord at $T=77$ K
for $d=\infty$ (a) and
$d=6$ nm (b); solid red line - $|a-b|=0.3$, dotted pink- $|a-b|=0.15$, and 
dashed blue - $|a-b|=0$.  }
\end{figure}

The upper and lower bounds often coincide, yielding the true value of the geometric discord.
This is specifically the case for pure states, Bell diagonal states, and states with vanishing
local Bloch vectors, $|x\rangle=|y\rangle=0$ \cite{miranowicz12}.
Hence, it is straightforward to show that the geometric discord is equal to $1/2$ for all
maximally entangled two-qubit states \cite{blanchard01},
\begin{equation}
\label{ini}
|\psi\rangle =
\sqrt{a}|00\rangle +\sqrt{b}e^{i\alpha}|10\rangle 
+\sqrt{b}e^{i\beta}|01\rangle -\sqrt{a}e^{i(\alpha+\beta)}|11\rangle),
\end{equation}
with $2a+2b=1$.
Furthermore, the discord of the Bell diagonal states ($a=0$ or $b=0$)
under phonon-induced partial pure dephasing is equal to 
$D_{S}(t)=2|\rho_{ij}(t)|^2=1/2\exp[2B_{ij}(t)]$, 
where $ij=12$ for $a=0$ and $03$ for $b=0$, and the appropriate forms of $B_{ij}(t)$
are given by Eqs. (\ref{B12}) and (\ref{B03}), which, up to a normalization, yields the
square of the concurrence.
\begin{figure}[th]
\begin{center}
\unitlength 1mm
\begin{picture}(85,60)(0,5)
\put(0,0){\resizebox{85mm}{!}{\includegraphics{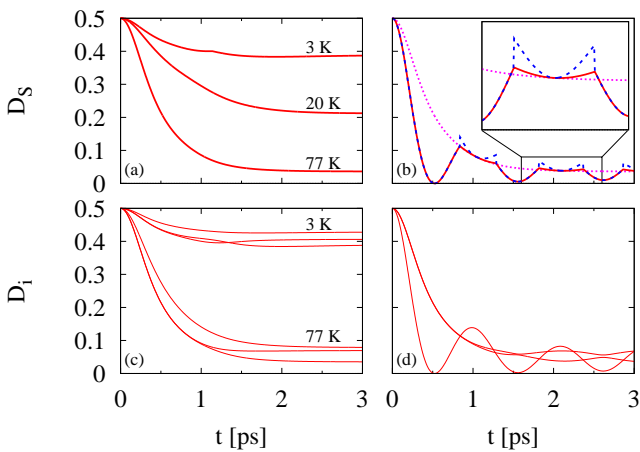}}}
\end{picture}
\end{center}
\caption{\label{evol4} Evolution of geometric discord bounds at $d=6$ nm for pure initial state (\ref{ini})
with $a=1/\sqrt{2}$ (a,b) at different temperatures for $\Delta E = 0$ for which the upper and lower 
bounds are equal (a) and at $77$ K
for $\Delta E = 6$ ps$^{-1}$; red solid line - lower bound, blue dashed - upper bound, pink dotted -
$\Delta E = 0$ (b). Lower bound values corresponding to the three eigenvalues of the matrix $K_x$
(the minimum of which yields the geometric discord lower bound) 
at different temperatures for zero biexcitonic shift (c) and and at $77$ K
for $\Delta E = 6$ ps$^{-1}$ (d).}
\end{figure}

Let us first study the evolution of the mixed X-state,
\begin{equation}
\label{X}
\rho=\left(
\begin{array}{cccc}
a&0&0&ag_{03}(t)\\
0&b&bg_{12}(t)&0\\
0&bg_{12}^*(t)&b&0\\
ag_{03}^*(t)&0&0&a
\end{array}
\right),
\end{equation}
which is significantly simpler (although hardly accessible experimentally), 
but already carries some of the properties 
of the discord evolution of a pure initial state with all coherences
present. The entanglement of such a state, measured by the concurrence, is equal to
$C(t)=\max\{0,b|g_{12}(t)|-a,a|g_{03}(t)|-b\}$ and is prone to sudden death.
The geometric discord is given by $D_S(t)= (a|g_{03}(t)|-b|g_{12}(t)|)^2+(a-b)^2$, if 
$|a-b|<a|g_{03}(t)|+b|g_{12}(t)|$ and by $D_S(t)= 2a^2|g_{03}(t)|^2 +2b^2|g_{12}(t)|^2$, if 
$|a-b|>a|g_{03}(t)|+b|g_{12}(t)|$ (for long times, if $a\neq b$). Hence, the discord will not undergo
sudden-death-like behavior, but, if $a\neq b$, it will display a transition between two types of decay
(there is no simple relation between the transition point and the point of entanglement
sudden death). The transition point coincides with the transition point between quantum 
and classical decoherence indicated in Ref. \cite{mazzola10}.
This is illustrated in Fig.~\ref{zgrzyt_dziwny2}, where the geometric discord
of the state (\ref{X}), with $g_{ij}(t)=\exp(-iA_{ij}(t)+B_{ij}(t))$, is plotted as a function 
of time for different values of $|a-b|$. The left panel corresponds to infinitely distant dots,
for which $g_{12}(t)=g_{03}(t)$, and the transition which is induced by the smooth partial
pure dephasing process is clearly visible. On the right panel, a similar evolution of the dots
separated by the distance $d=6$ nm is shown, which additionally displays an enhancement
of the geometric discord after a finite time. This effect is due to a positive interference
between phonon wave packets traveling from the two dots. Note, that the process is sufficient
to induce quantum correlations in an initially uncorrelated state with $a=b$ (which remains
uncorrelated, if $d=\infty$).

The next step is to study the evolution of the lower and upper geometric discord bounds
for an initial state (\ref{ini}) with all non-zero coherences ($a\neq 0$ and $b\neq 0$)
under phonon-induced partial pure dephasing. For simplicity
the studied state is taken with $a=b=1/4$ (the local phases $\alpha$ and $\beta$
do not change the values of the geometric discord or either of its bounds).
On Fig. \ref{evol4} (a) the evolutions of the geometric discord are plotted at different
temperatures for zero biexcitonic shift (the upper and lower bounds
are equal in this case). The $3$ K curve shows a distinct point where the discord is not smooth,
resembling the evolution of the X-state (\ref{X}),
which is absent at higher temperatures. To understand this, the evolutions of
$D_i=\Tr[K_x]-k_i$, where $k_i$ are the three eigenvalues of the matrix $K_x$
(the minimum of $D_i$ yields the true lower bound of the geometric discord) for $3$ and $77$ K
are plotted in Fig. \ref{evol4} (c). At $3$ K a crossing of two $D_i$ curves is observed which
is caused by the positive interference of phonon wave packets, which is responsible
for the enhancement of the geometric discord for the X-state of eq. (\ref{X}).

Fig. \ref{evol4} (b) shows the evolution of the lower (red solid line) and upper (blue dashed line)
bounds on the geometric discord for the same initial state at $77$ K when the biexcitonic
shift is nonzero. The biexcitonic shift in the absence of any decoherence processes causes
a coherent oscillation between the initial, maximally entangled state, and the separable state
$|\psi_{\mathrm{sep}\rangle} = 1/2(|0\rangle+|1\rangle )\otimes(|0\rangle+|1\rangle)$
(reached when $\Delta E t =(2n+1)\pi$, where $n$ is a natural number).
Under phonon-induced pure dephasing, the oscillations of entanglement are damped and display prolonged 
periods when the entanglement is zero (which is only possible when the damping process
can lead to sudden death of entanglement), and are otherwise smooth while their amplitude
is limited by the entanglement decay displayed by the zero-biexcitonic shift evolution \cite{roszak06a}.
The oscillations of the geometric discord, which without decoherence would mimic entanglement oscillations,
are substantially different. Firstly, the discord does not display sudden-death-type behavior,
and approaches states with $\Delta E t =(2n+1)\pi$ smoothly, reaching zero at $t =(2n+1)\pi/\Delta E $
only, if inter-phonon-interference does not induce extra coherence in the system
(at short times and/or long distances between the dots). This confirms the notion that,
since the set of zero-discord (only classically correlated) states has zero volume, decoherence 
processes will never lead to the sudden and permanent vanishing of the quantum discord \cite{ferraro10}.

Furthermore, the evolution of the quantum discord induced by the biexcitonic shift leads
to the situation, when the value of the geometric discord is greater than the corresponding
zero-biexcitonic-shift value.This can be clearly seen in the inset of Fig.~\ref{evol4} (b),
where both the lower and upper bound of the geometric discord exceed the zero-biexcitonic-shift curve
(pink dotted line). This behavior is non-monotonous and symmetric (for constant decoherence) 
with respect to the 
maximally entangled points given by $\Delta E t =2n\pi$ (for which the non-zero-biexcitonic-shift
and zero-biexcitonic-shift lines have to coincide).
This shows that the dependence of the discord on quantum phase relations is non-trivial,
and than non-local phase correlations may lead to an enhancement of quantum correlations
in mixed states
depending on the actual value of the phase factor.
The fact that the lower and upper bounds on the geometric discord are different in this case
is in agreement with predictions made in Ref.~\cite{miranowicz12}. 
We surmise that the discontinuity of the upper bound and the sharp features of the lower
bound of the discord are an artifact of the procedure of their generation from the density
matrix, while the
actual curve of the geometric discord is continuous and smooth.
Fig. \ref{evol4} (d) shows the evolutions of the lower bound values, $D_i$, corresponding
to the three eigenvalues of the matrix $K_x$, the minimum of which yields the actual lower bound, 
to illustrate the origin of the irregular shape of the lower bound of the geometric discord.

We have studied the evolution of the geometric discord of a two QD qubit system under
decoherence caused by the phonon environment, giving the lower and upper bounds on the discord
where it was impossible to find its true value. We have shown that the discord does not display
sudden death type behavior, but reveals a number of characteristic features  
(which are not displayed by entanglement) under 
the influence of phonons, 
which cause a continuous and smooth partial pure dephasing process.
Firstly, the evolution of the geometric discord often displays a transition between
different types of decay, which is particularly evident for initial entangled X-states,
but has also been observed for maximally entangled pure states with all coherences present.
The study of the evolution of the discord in these pure initial states
showed the importance of non-local phase correlations; a shift in the phase can lead to 
the enhancement of quantum correlations in a mixed two-qubit state.
Furthermore, the positive interference of phonon wave packets originating from the two dots
(interaction through a common reservoir)
which is weak in the system and 
cannot generate entanglement between separable states, does lead to the appearance
of quantum correlations described by the discord.
Hence, the study of the quantum discord in this realistic scenario shows, among other things, that 
quantum correlations are a common occurrence in mixed separable states.

\begin{acknowledgments}
This work was supported by the TEAM programme of the Foundation for Polish Science co-financed from the European Regional Development Fund (K. R.).
P.M. was supported by the Foundation for Polish Science International PhD Projects Programme co-financed by the EU European Regional Development Fund.
P. H. acknowleges support from the
National Science Centre project Maestro DEC-2011/02/A/ST2/00305.
\end{acknowledgments}

\end{document}